\documentclass[a4paper,10pt,twoside]{cpc-hepnp}

\usepackage{multicol}
\usepackage{graphicx}
\usepackage{booktabs}
\usepackage{amssymb,bm,mathrsfs,bbm,amscd}
\usepackage[tbtags]{amsmath}
\usepackage{lastpage}

\begin{document}

\fancyhead[co]{\footnotesize V.V. Anashin et al: Measurement of $J/\psi\to\eta_{\rm c}\gamma$ at KEDR}

\footnotetext[0]{Received ?? December 2009}

\title{Measurement of $J/\psi\to\eta_{\rm c}\gamma$ at KEDR\thanks{Partially supported by 
the Russian Foundation for Basic Research, grants 08-02-00258, 08-02-00258, and RF 
Presidential Grant for Sc. Sch. NSh-5655.2008.2.}
}

\author{
V.V. Anashin$^1$ V.M. Aulchenko$^{1,2}$ E.M. Baldin$^{1,2}$ A.K. Barladyan$^1$ A.Yu. Barnyakov$^1$ M.Yu. Barnyakov$^1$ \\
S.E. Baru$^{1,2}$ I.V. Bedny$^1$ O.L. Beloborodova$^{1,2}$ A.E. Blinov$^1$ V.E. Blinov$^{1,3}$ A.V. Bobrov$^1$ \\
V.S. Bobrovnikov$^1$ A.V. Bogomyagkov$^{1,2}$ A.E. Bondar$^{1,2}$ A.R. Buzykaev$^1$ S.I. Eidelman$^{1,2}$ \\
Yu.M. Glukhovchenko$^1$ V.V. Gulevich$^1$ D.V. Gusev$^1$, S.E. Karnaev$^1$ G.V. Karpov$^1$ S.V. Karpov$^1$ \\
T.A. Kharlamova$^{1,2}$ V.A. Kiselev$^1$ S.A. Kononov$^{1,2}$ K.Yu. Kotov$^1$ E.A. Kravchenko$^{1,2}$ \\
V.F. Kulikov$^{1,2}$ G.Ya. Kurkin$^{1,3}$ E.A. Kuper$^{1,2}$ E.B. Levichev$^{1,3}$ D.A. Maksimov$^1$ %
V.M. Malyshev$^{1;1)}$\email{V.M.Malyshev@inp.nsk.su} %
\\
A.L. Maslennikov$^1$ A.S. Medvedko$^{1,2}$  O.I. Meshkov$^{1,2}$ A.I. Milstein$^1$ S.I. Mishnev$^1$ \\
I.I. Morozov$^{1,2}$ N.Yu. Muchnoi$^{1,2}$  V.V. Neufeld$^1$ S.A. Nikitin$^1$ I.B. Nikolaev$^{1,2}$ I.N. Okunev$^1$ \\
A.P. Onuchin$^{1,3}$ S.B. Oreshkin$^1$ I.O. Orlov$^{1,2}$ A.A. Osipov$^1$ S.V. Peleganchuk$^1$ S.G. Pivovarov$^{1,3}$ \\
P.A. Piminov$^1$ V.V. Petrov$^1$ A.O. Poluektov$^1$ I.N. Popkov$^1$ G.E. Pospelov$^1$ V.G. Prisekin$^1$ A.A. Ruban$^1$ \\
V.K. Sandyrev$^1$ G.A. Savinov$^1$ A.G. Shamov$^1$ D.N. Shatilov$^1$ B.A. Shwartz$^{1,2}$ E.A. Simonov$^1$ \\
S.V. Sinyatkin$^1$ Yu.I. Skovpen$^{1,2}$ A.N. Skrinsky$^1$ V.V. Smaluk$^{1,2}$ A.V. Sokolov$^1$ A.M. Sukharev$^1$ \\
E.V. Starostina$^{1,2}$ A.A. Talyshev$^{1,2}$ V.A. Tayursky$^1$ V.I. Telnov$^{1,2}$ Yu.A. Tikhonov$^{1,2}$ \\
K.Yu. Todyshev$^{1,2}$ G.M. Tumaikin$^1$ Yu.V. Usov$^1$ A.I. Vorobiov$^1$ A.N. Yushkov$^1$ V.N. Zhilich$^1$ \\
V.V.Zhulanov$^{1,2}$  A.N. Zhuravlev$^{1,2}$ \\
}
\maketitle

\address{%
1~(Budker Institute of Nuclear Physics, 11, Lavrentiev prospect, Novosibirsk, 630090, Russia)\\
2~(Novosibirsk State University, 2, Pirogova street, Novosibirsk, 630090, Russia)\\
3~(Novosibirsk State Technical University, 20, Karl Marx prospect, Novosibirsk, 630092, Russia)
}

\begin{abstract}
We present a study of the inclusive photon spectra from 5.9 million $J/\psi$ decays collected with the KEDR detector
at the VEPP-4M $e^+e^-$ collider.
We measure the branching 
fraction of radiative decay $J/\psi\to\eta_{\rm c}\gamma$, $\eta_{\rm c}$ width and mass. Our preliminary results are: 
$M(\eta_{\rm c}) = 2979.4 \pm 1.5 \pm 1.9$ MeV/$c^2$, $\Gamma(\eta_{\rm c}) = 27.8 \pm 5.1 \pm 3.3$ MeV, 
${\cal B}(J/\psi\to\eta_{\rm c}\gamma) = (2.34\pm0.15\pm0.40)\%$.
\end{abstract}

\begin{keyword}
$\eta_{\rm c}$ meson, full width, mass, radiative decay
\end{keyword}

\begin{pacs}
 13.20.Gd, 13.40.Hq, 14.40.Pq
\end{pacs}

\begin{multicols}{2}

\section{Introduction}
$J/\psi\to\eta_{\rm c}\gamma$ decay is an M1 radiative transition
with photon energy about 114 MeV, and a relatively large branching fraction about 2\%.
This is a transition between 1S states of charmonium, which rate can be easily calculated in potential models. 
However, for a long time a significant difference between theoretical predictions and experimental data was observed.
Until the last year the PDG value of this branching was based on a single measurement\cite{lab1} of Crystal Ball using 
an inclusive spectrum of photons in the $J/\psi$ decays
with the result ${\cal B}(J/\psi\to\eta_{\rm c}\gamma)=(1.27\pm0.36)\%$, approximately twice less than the theoretical 
predictions $(2.9\pm0.5)\%$\cite{lab2}, $(2.0\pm0.8)\%$\cite{lab3}, $(2.4\pm0.3)\%$\cite{lab4}, 
$(1.6\pm1.0)\%$\cite{lab5}, $(2.1\pm0.4)\%$\cite{lab6}, $(3.05\pm0.07)\%$\cite{lab7}. For comparison, 
in E1 transitions of charmonium the corresponding difference is (20-30)\% only\cite{lab7}, although the matrix 
element has more complex structure.

In 2009 CLEO published\cite{lab8} a new measurement of this branching with the value 
${\cal B}(J/\psi\to\eta_{\rm c}\gamma)=(1.98\pm0.09\pm0.30)\%$, which is closer to theory. CLEO analysis used 
exclusive $\eta_{\rm c}$ decays and found that the line shape of this transition 
is asymmetric. A Breit-Wigner function (BW) alone provides a poor fit to data. 
A Breit-Wigner modified by $E_{\gamma}^3$ 
(according to theoretical predictions for electromagnetic dipole transitions)
improves the fit near a resonance, but gives a diverging tail at higher photon energies. 
To suppress this tail CLEO used an additional exponential factor.

\section{Photon spectrum}
The photon spectrum in $J/\psi \to \eta_{\rm c}\gamma$ decay is given by the formula
\begin{equation}
\label{fone}
\frac{dN_{\gamma}}{d\omega} = N_{\psi}{\cal B} \vint\nolimits_{0}^{M_{\psi}/2} d\omega'
\frac{d\Gamma(\omega')}{d\omega'}\frac{\epsilon(\omega')g(\omega,\omega')}{\Gamma_{\eta_{\rm c}\gamma}},
\end{equation}
where $\cal B$ is the decay branching fraction, $g(\omega,\omega')$ is a calorimeter response function, 
$\epsilon(\omega)$ is a photon detection efficiency, $d\Gamma(\omega)/d\omega$ is a true photon spectrum, 
and $\Gamma_{\eta_{\rm c}\gamma}=\vint\nolimits d\omega d\Gamma(\omega)/d\omega$ 
is the full decay width.
The true photon spectrum can be written as \cite{lab2}
\begin{equation}
\frac{d\Gamma(\omega)}{d\omega} = \frac{4}{3}\alpha \frac{e_{\rm c}^2}{m_{\rm c}^2}\omega^3|M|^2BW(\omega).
\end{equation}
Here $M=<\eta_{\rm c}|j_0(\omega r/2)|J/\psi>$ is the matrix element of the transition, $j_0(x)=sin(x)/x$, $e_{\rm c}$ 
and $m_{\rm c}$ are c-quark charge and mass. 

We see that if $\omega$ tends to zero, the matrix element tends to unity.
CLEO used in their fit $|M|^2=exp(-\omega^2/8\beta^2)$ with $\beta=65$ MeV, but such a form of matrix element squared is valid for 
harmonic oscillator wave functions only. 
In all other potentials $|M|^2$ dependence will be proportional to $\omega^{-n}$. 

Therefore we tried to fit the CLEO data using another line shape: at photon energy
 $\omega$ near resonance the decay probability $d\Gamma/d\omega$ is proportional to $\omega^3$, 
 but at higher energies the factor $\omega^3$ is replaced with $\omega$. 
We found that the function $d\Gamma/d\omega\sim \omega^3\omega_0^2/(\omega\omega_0+(\omega-\omega_0)^2))BW(\omega)$,
 where $\omega_0$ is the most probable transition energy, is also suitable. 
In Fig.~\ref{fig1}, our fit of CLEO data using such a function is shown. 
\begin{center}
\includegraphics[width=\columnwidth]{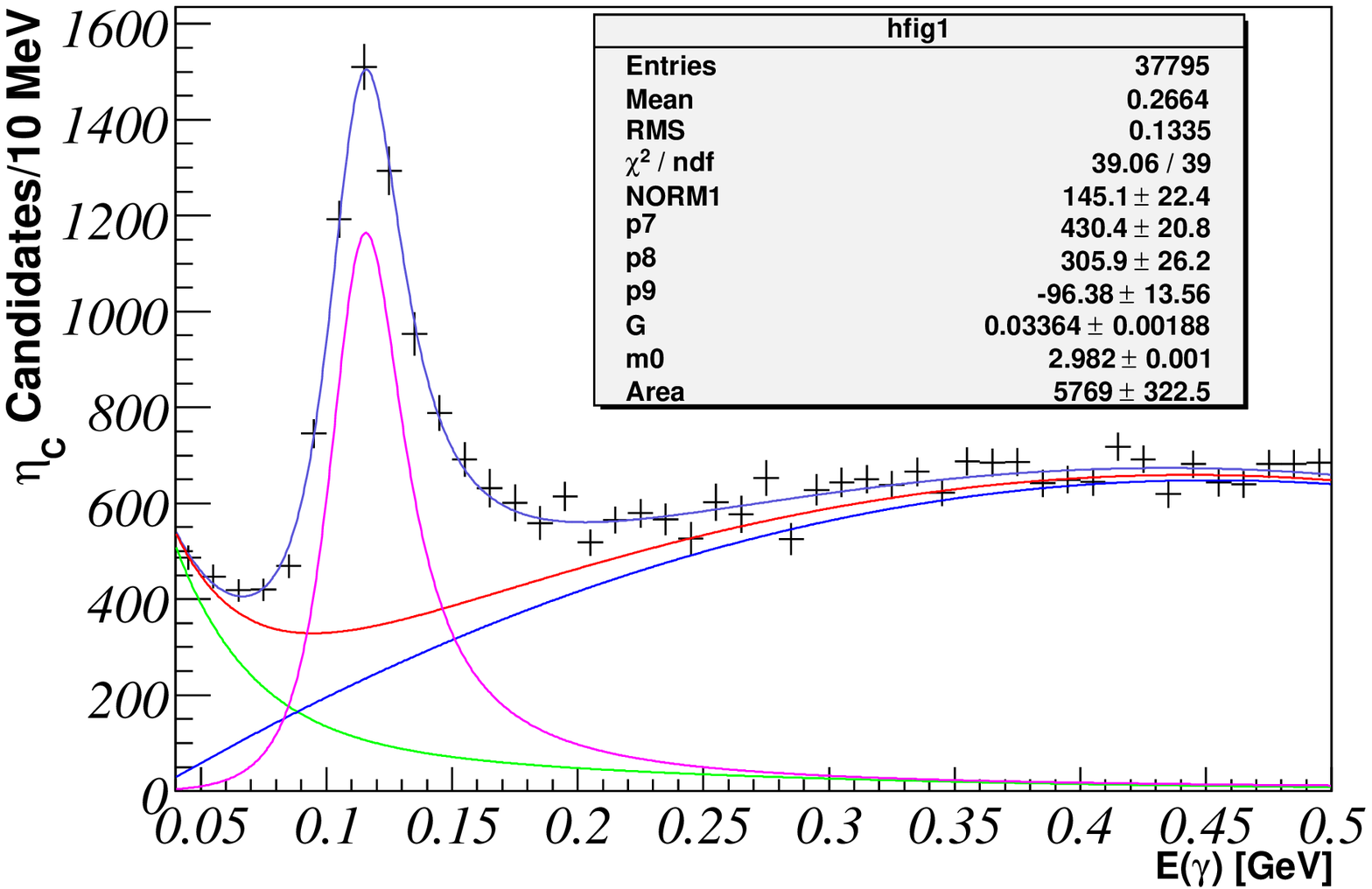}
\figcaption{\label{fig1}Fit of CLEO data using $d\Gamma/d\omega\sim  \omega^3\omega_0^2/(\omega\omega_0+(\omega-\omega_0)^2)BW(\omega)$.}
\end{center}
In Table~\ref{tab1} results of fits with the CLEO function and our function are shown. We see that confidence levels are good in both cases.

\end{multicols}
\ruleup
\begin{center}
\tabcaption{ \label{tab1}  Fit results of CLEO data using various decay probability functions.}
\vspace{-3mm} \footnotesize
\begin{tabular*}{170mm}{@{\extracolsep{\fill}}ccccc}
\toprule $d\Gamma/d\omega$ & $M(\eta_{\rm c}), MeV/c^2$ & $\Gamma(\eta_{\rm c}), MeV$  & $N_{1S}^{EXC}$ & $\chi^2/NDF (C.L.)$ \\
\hline
$\sim\omega^3exp(-\omega^2/8\beta^2)BW(\omega)$ & $2982.4\pm0.7$ & $32.5\pm1.8$ & $6142\pm430$ & 38.0/38 (0.47) \\
$\sim\omega^3\omega_0^2/(\omega\omega_0+(\omega-\omega_0)^2)BW(\omega)$ & $2981.8\pm0.5$ & $33.6\pm1.9$ & $6494\pm362$ & 39.1/39 (0.47) \\
\bottomrule
\end{tabular*}%
\end{center}
\vspace{0.5cm}
\ruledown 

\begin{multicols}{2}

\section{KEDR data}
The experiment was performed at the KEDR detector\cite{lab9} of the VEPP-4M collider\cite{lab10}. 
In Fig.~\ref{fig2} a layout of the VEPP-4M collider is shown. 
It operates at a peak luminosity about $1.5\cdot10^{30}$ ${\rm cm}^{-2}{\rm s}^{-1}$ near the $J/\psi$ peak energy. 
Two methods of beam energy determination are used: resonant depolarization with accuracy of
$8\div30$ keV and IR-light Compton backscattering with accuracy $\sim100$ keV\cite{lab11}.

The view of the KEDR detector is shown in Fig.~\ref{fig3}.  
Subsystems are listed in the figure. 
In this analysis a barrel liquid krypton calorimeter, drift chamber (DC) and time-of-flight (ToF) scintillator counters 
were of decisive importance. 
Luminosity was measured by single Bremsstrahlung online and by small angle Bhabha scattering offline.

This analysis is based on a data sample of $1.52\pm0.08$ ${\rm pb}^{-1}$ collected at the $J/\psi$ peak.
Three $J/\psi$ scans were performed; 
a measured beam energy spread was used for the calculation of the number of $J/\psi$ produced: $N_{\psi}=(5.9\pm0.3)\cdot10^6$.

Event selection was done in two steps. At the first step multihadron decays of $J/\psi$ were selected. To do this the 
following cuts were applied: 
total energy in clusters in calorimeters is greater than 0.8 GeV; at least 4 clusters with energy greater than 30 MeV in 
calorimeters are detected; 
at least one central track in the drift chamber is reconstructed; there are no muon tubes activated in the third layer. These cuts effectively suppress 
background from the cosmic rays, beam-gas interactions and Bhabha events. 

\begin{center}
\includegraphics[width=\columnwidth]{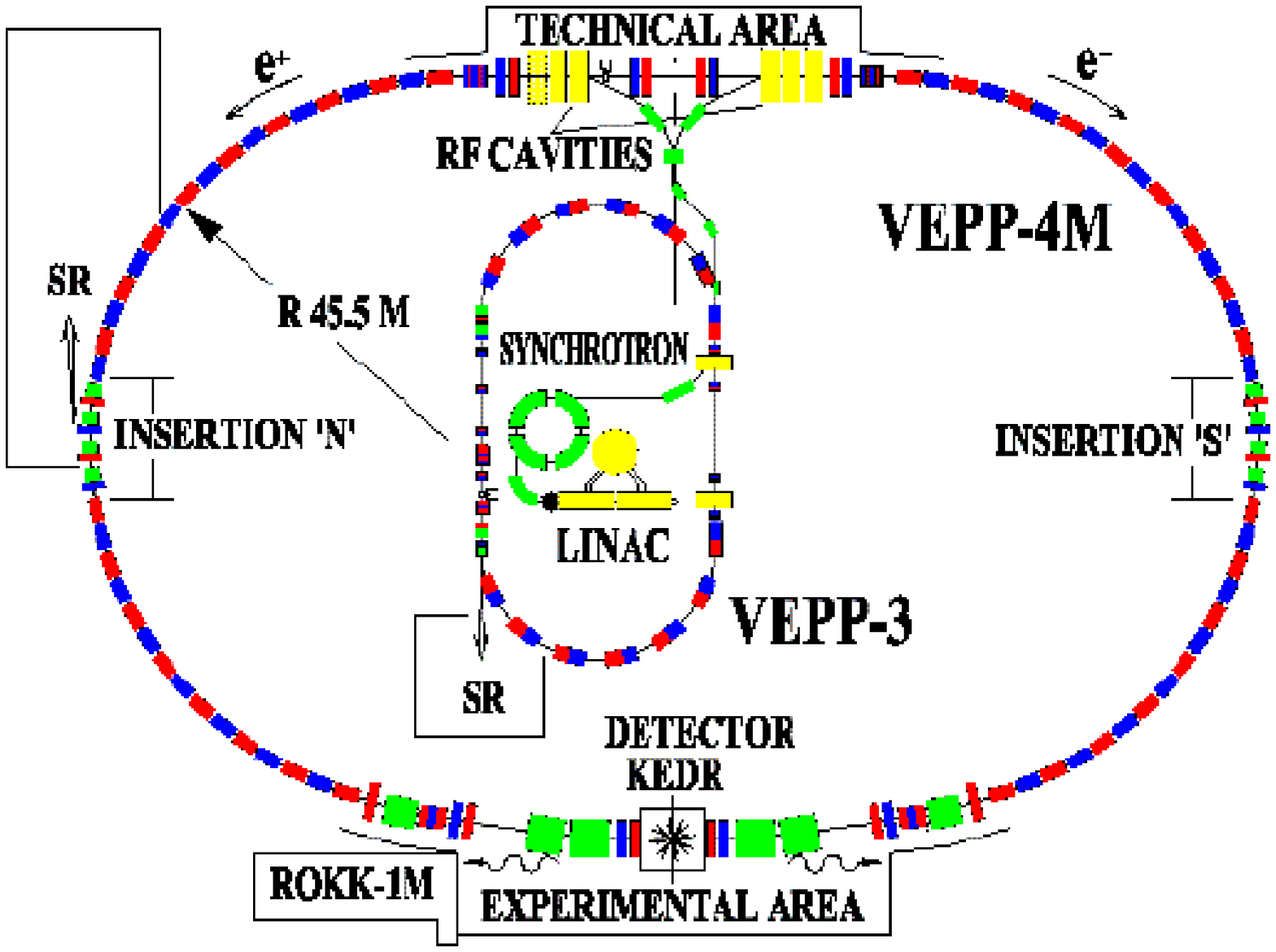}
\figcaption{\label{fig2}VEPP-4M collider.}
\end{center}

\begin{center}
\includegraphics[width=\columnwidth]{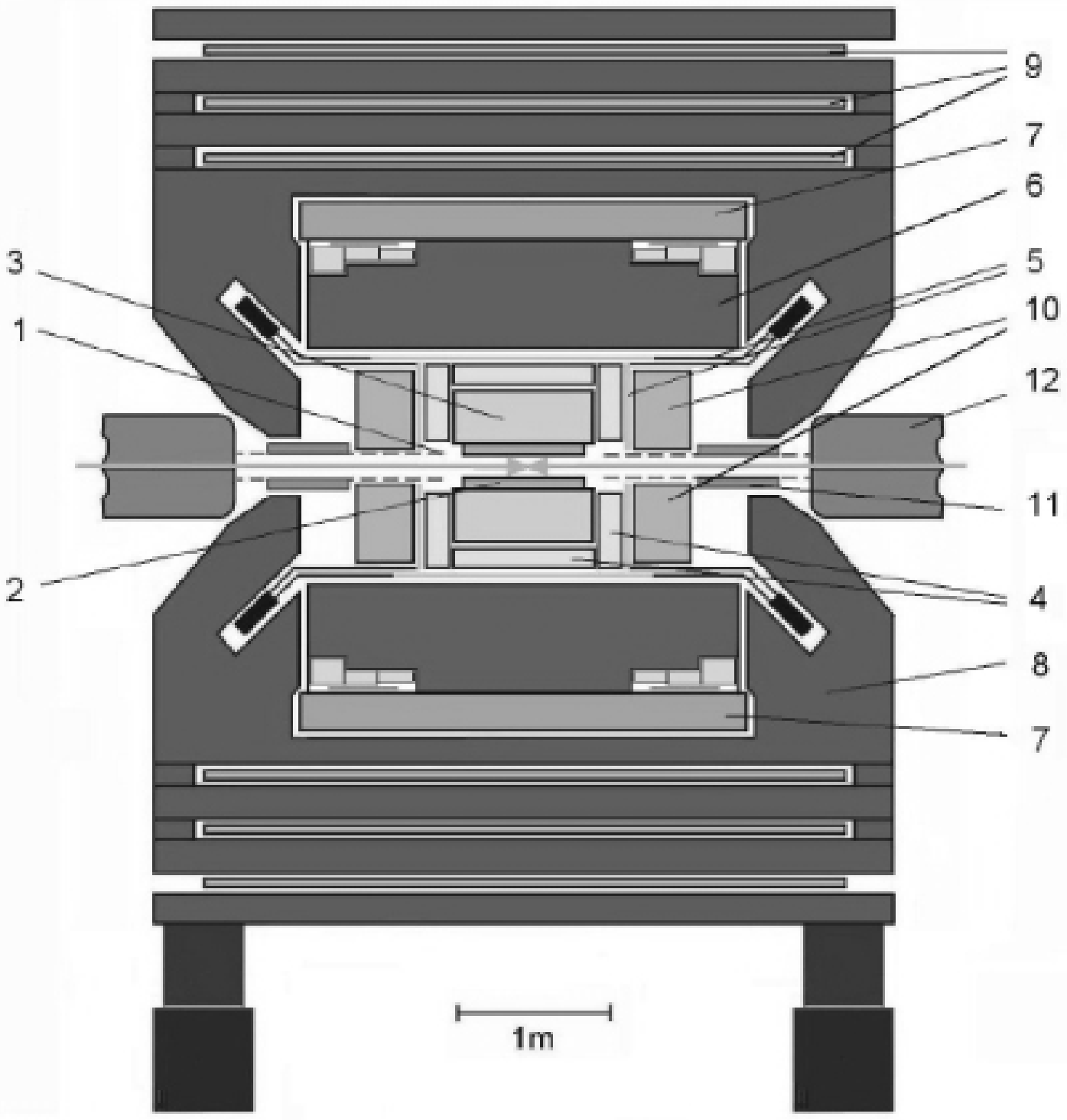}
\figcaption{\label{fig3}KEDR detector. 1-vacuum chamber, 2-vertex detector, 3-drift chamber, 4-threshold aerogel counters, 
5-ToF-counters, 6-liquid krypton calorimeter, 7-superconducting coil (0.65 T), 8-magnet yoke, 9-muon tubes, 10-CsI-calorimeter,
11-compensation solenoid, 12-VEPP-4M quadrupole.
}
\end{center}

At the second step photons in these events were identified. 
A photon is a cluster in the liquid krypton calorimeter without reconstructed tracks in the drift chamber attached to it 
and without ToF scintillator counters activated before the cluster. According to the simulation, the photon detection 
efficiency with the above mentioned cuts is about 34\%.

\section{Data analysis}
The inclusive spectra of clusters are shown in Fig.~\ref{fig4}. 
Large peaks in the spectra correspond to minimal ionizing
particles crossing the calorimeter.
The spectrum of charged particles was used to reject charged particles 
which were detected as neutrals from the photon spectrum during fitting procedure.
\begin{center}
\includegraphics[width=\columnwidth]{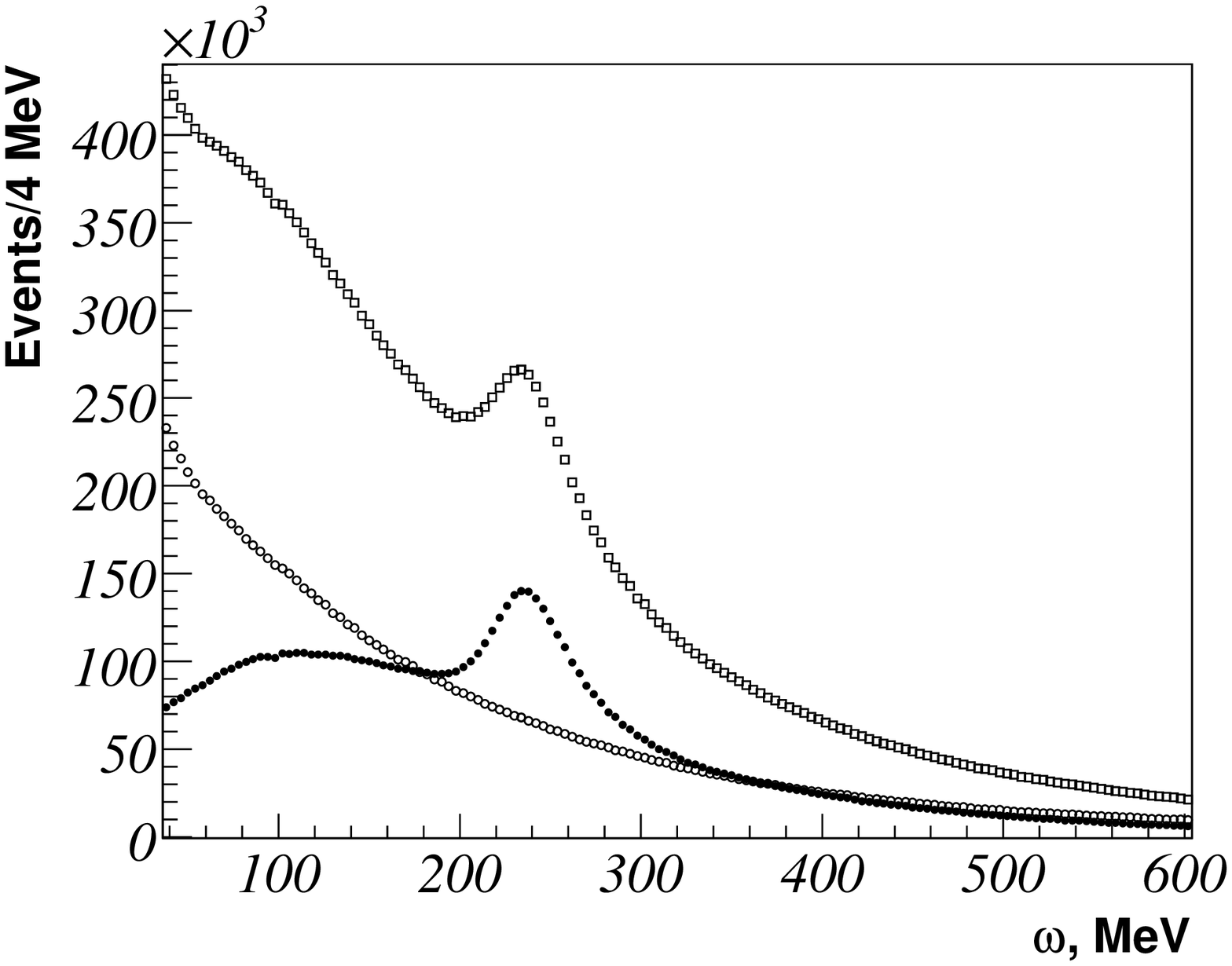}
\figcaption{\label{fig4}Inclusive spectra in the liquid krypton calorimeter: 
 open squares - all clusters; shaded circles - charged particles, 
 when the ToF scintillator counter is activated before a cluster and a DC track is attached to cluster;
 open circles - photons.
}
\end{center}
In Fig.~\ref{fig5} a fit to our data is shown. 
\begin{center}
\includegraphics[width=\columnwidth]{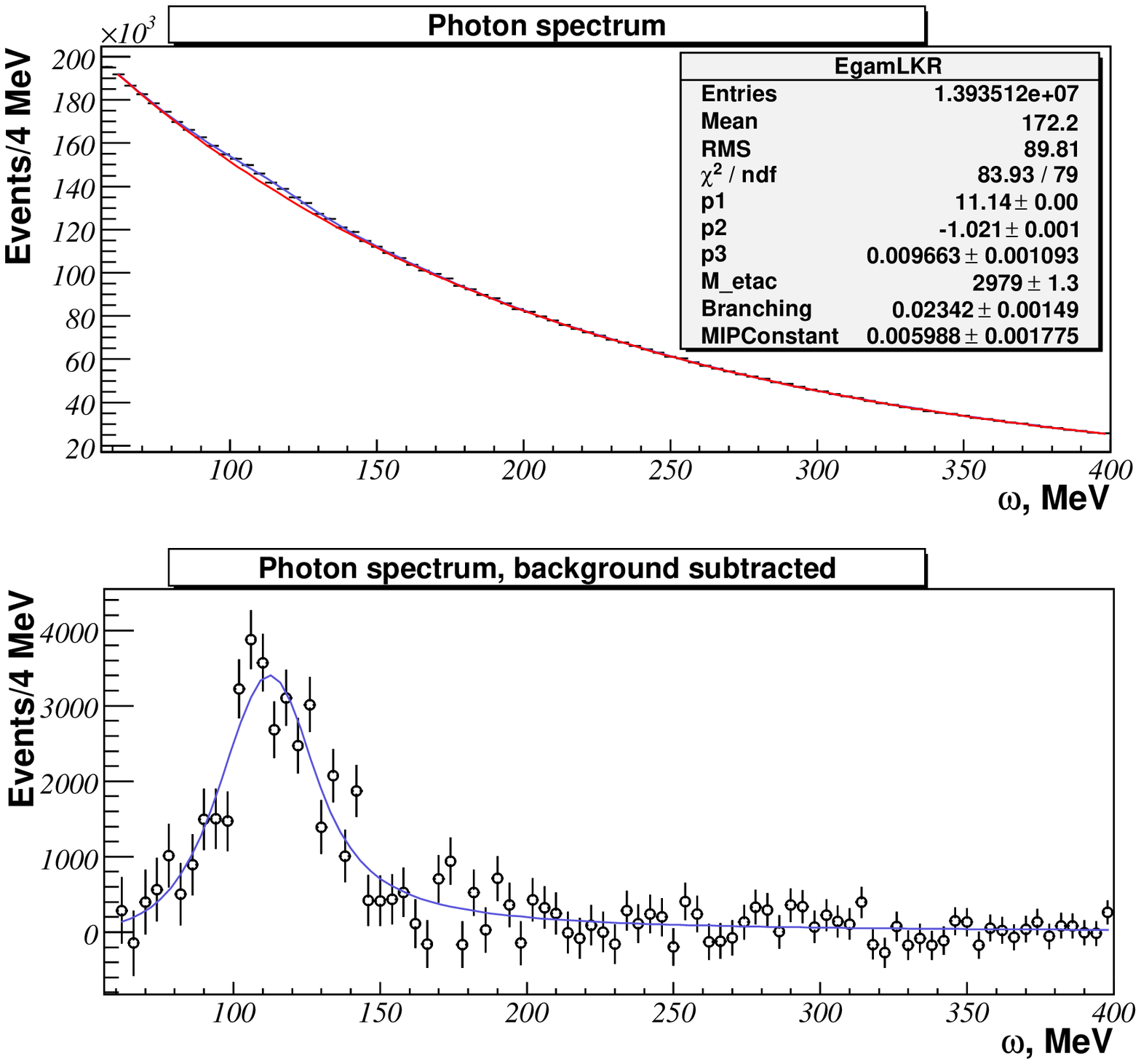}
\figcaption{\label{fig5} Fit of the inclusive photon spectrum with 
$f(\omega) = \omega_0^2/(\omega\omega_0+(\omega-\omega_0)^2)$. 
}
\end{center}
The inclusive photon spectrum was fit by the sum of the signal having a shape 
$d\Gamma/d\omega \sim \omega^3 f(\omega)BW(\omega)$, convolved with the calorimeter response function 
(Novosibirsk PDF - asymmetric Gaussian with $\sigma_E=8.8$ MeV at 110 MeV and a=-0.3), and background. 
\begin{center}
\includegraphics[width=\columnwidth]{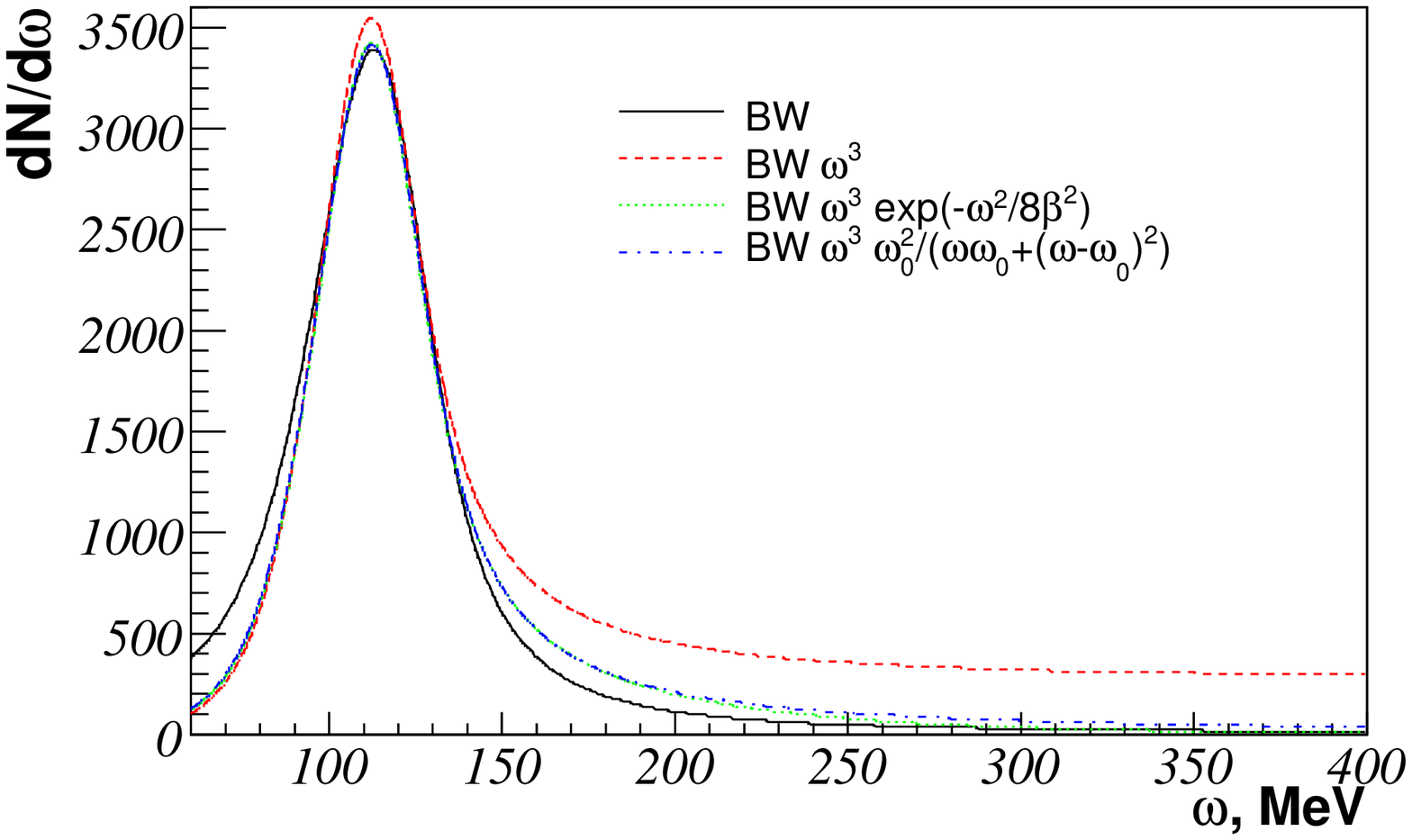}
\figcaption{\label{fig6} Resulting signal line shapes for various $d\Gamma/d\omega$ functions.
}
\end{center}
Background was taken in the form of exponent of a second-order polynomial plus a scaled histogram of 
charged particles. A relativistic Breit-Wigner function was used in the fit.

We also tried to fit our data with various line shapes. We used $BW$ alone, $BW\cdot \omega^3$, CLEO function, and our function. 
The resulting signal line shapes for these functions are shown in Fig.~\ref{fig6}.
The BW function alone gives a shifted value of $\eta_{\rm c}$ mass compared to other functions: 
$M(\eta_{\rm c})=2975.9\pm1.3$ MeV/$c^2$. 
$BW\cdot\omega^3$ gives a better form near the resonance, 
but leads to the unphysical diverging tail at higher photon energies. 
The last two functions give close results of fits, and their difference can be used for estimation of a systematic 
error appearing due to the unknown line shape.
In Table~\ref{tab2} results of fits of KEDR data with the CLEO function and our function are listed. 
We see that confidence levels are good in both cases.

\end{multicols}
\ruleup
\begin{center}
\tabcaption{ \label{tab2}  Fit results of KEDR data using various decay probability functions.}
\vspace{-3mm} \footnotesize
\begin{tabular*}{170mm}{@{\extracolsep{\fill}}ccccc}
\toprule $d\Gamma/d\omega$ & $M(\eta_{\rm c}), MeV/c^2$ & $\Gamma(\eta_{\rm c}), MeV$  & ${\cal B}(J/\psi)\to\gamma\eta_{\rm c}), \%$ & $\chi^2/NDF (C.L.)$ \\
\hline
$\sim\omega^3exp(-\omega^2/8\beta^2)BW(\omega),$ $\beta=65 MeV$ & $2979.7\pm1.6$ & $26.9\pm4.8$ & $2.19\pm0.28$ & 84.1/78 (0.30) \\
$\sim \omega^3\omega_0^2/(\omega\omega_0+(\omega-\omega_0)^2)BW(\omega)$ & $2979.4\pm1.5$ & $27.8\pm5.1$ & $2.36\pm0.34$ & 83.9/78 (0.30) \\
\bottomrule
\end{tabular*}%
\end{center}
\vspace{0.5cm}
\ruledown 

\begin{multicols}{2}

\section{Systematic errors}
In Table~\ref{tab3} systematic errors of our measurements are listed. 
Systematic error appearing due to the unknown line shape was estimated using various line shapes and 
fixing $\eta_{\rm c}$ width to the current PDG values ($27.4\pm2.9$ MeV).
Systematic error related to the background subtraction was estimated by the taking a polynomial of the third order instead 
of the second in the fit, and varying ranges of the fit.
Luminosity measurement error was evaluated from the difference between the measured and calculated number of the multihadron $J/\psi$ decays.
Photon detection efficiency is conservatively estimated as 10\%.
Photon energy scale calibration was made comparing a $\pi^0$ peak position in simulation and data.

\end{multicols}
\ruleup
\begin{center}
\tabcaption{ \label{tab3}  Systematic errors.}
\vspace{-3mm} \footnotesize
\begin{tabular*}{170mm}{@{\extracolsep{\fill}}lccc}
\toprule Systematic error & $M(\eta_{\rm c}), MeV/c^2$ & $\Gamma(\eta_{\rm c}), MeV$  & ${\cal B}(J/\psi)\to\gamma\eta_{\rm c}), \%$ \\
\hline
 Line shape & 1.5 & 2.5 & 0.25  \\
 Background subtraction & 0.5 & 2.2 & 0.18  \\
 Luminosity measurement &  &  & 0.11  \\
 Photon efficiency &  &  & 0.23  \\
 Photon energy scale & 1.1 &  &   \\
\hline
 Total & 1.9 & 3.3 & 0.40  \\
\bottomrule
\end{tabular*}%
\end{center}
\vspace{0.5cm}
\ruledown 

\begin{multicols}{2}

\section{Results and conclusions}
A new direct measurement of $J/\psi\to\eta_{\rm c}\gamma$ decay was performed. 
We measured $\eta_{\rm c}$ mass, width, and branching fraction 
of $J/\psi\to\eta_{\rm c}\gamma$ decay. Our preliminary results are: 

$M(\eta_{\rm c}) = 2979.4 \pm 1.5 \pm 1.9$ MeV/$c^2$,

$\Gamma(\eta_{\rm c}) = 27.8 \pm 5.1 \pm 3.3$ MeV,

${\cal B}(J/\psi\to\eta_{\rm c}\gamma) = (2.34\pm0.15\pm0.40)\%$.

In Fig.~\ref{fig7} results of our measurements, compared with the other experiments and the current PDG values, are shown. 
The branching fraction of this decay is sensitive to the line shape, which is asymmetric, and it should be taken into account 
during a measurement.

We see that our results on $\eta_{\rm c}$ mass and width are consistent with the current PDG values.
Our result for ${\cal B}(J/\psi\to\eta_{\rm c}\gamma)$ is consistent with that of CLEO, is higher than the old Crystal Ball value and close 
to theoretical predictions.

We are grateful to N.Brambilla, A.Yu. Khodjamirian, M.A. Shifman and A. Vairo for valuable comments.
\end{multicols}
\begin{center}
\includegraphics[width=5.5cm]{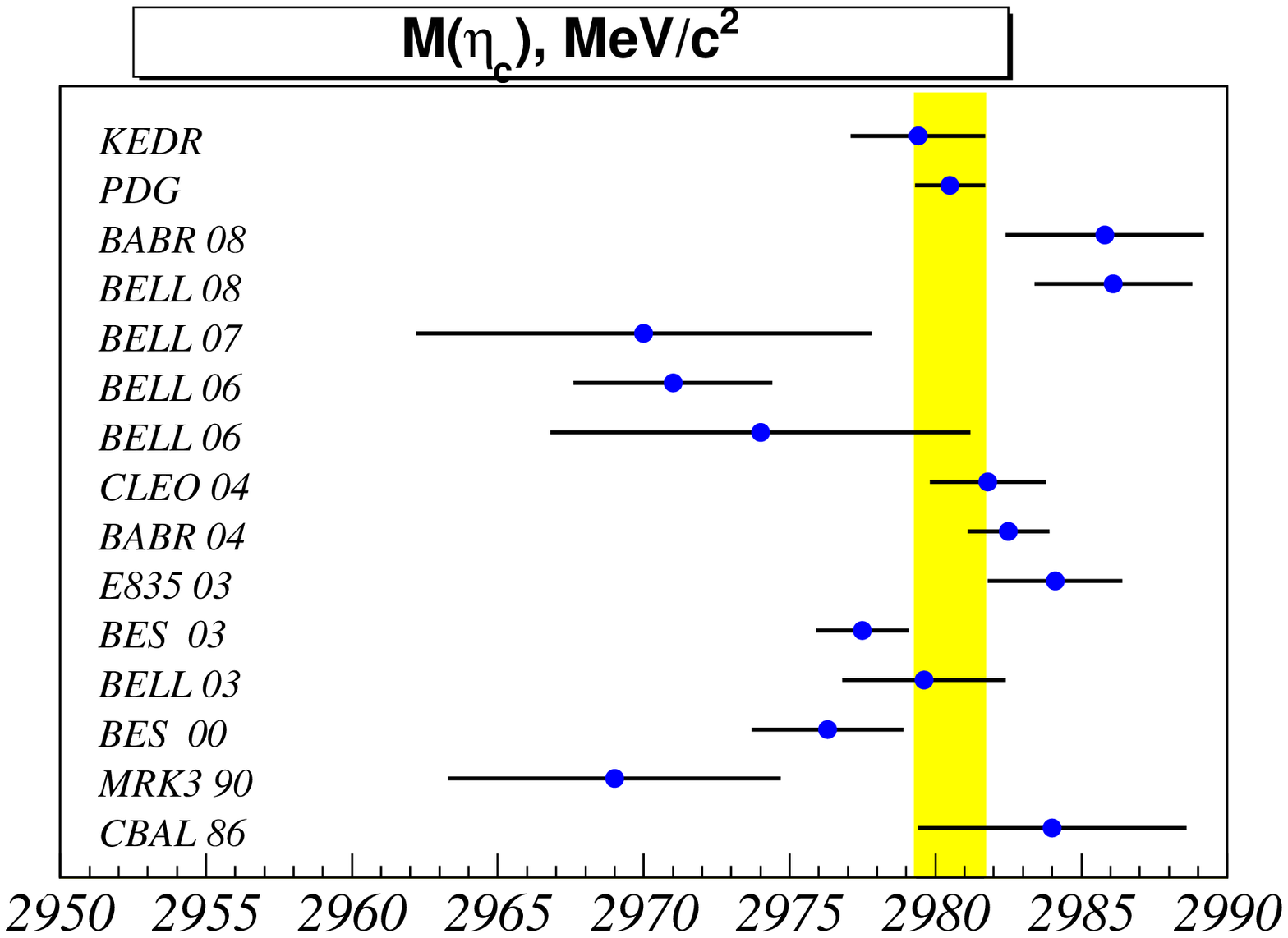}
\includegraphics[width=5.5cm]{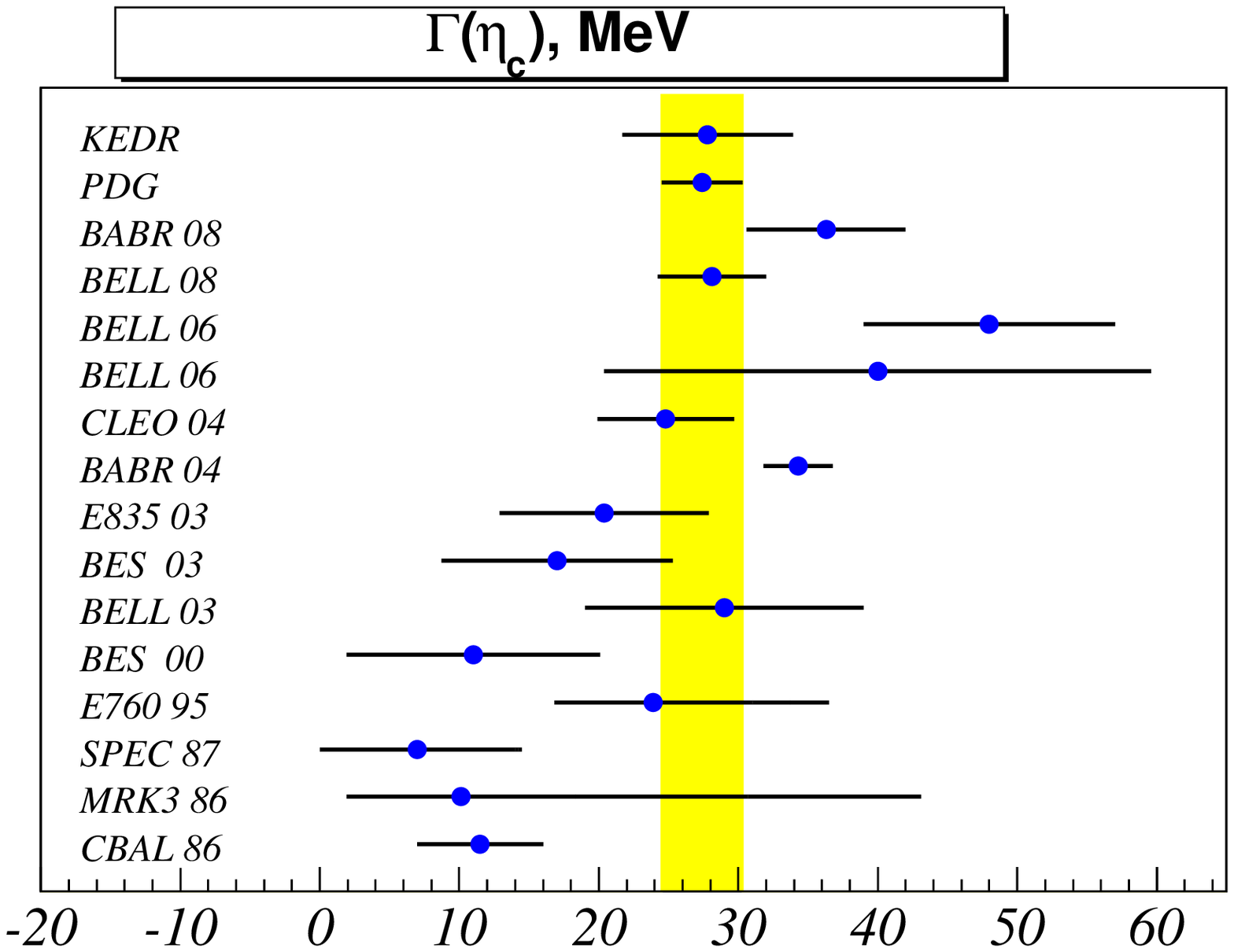}
\includegraphics[width=5.5cm]{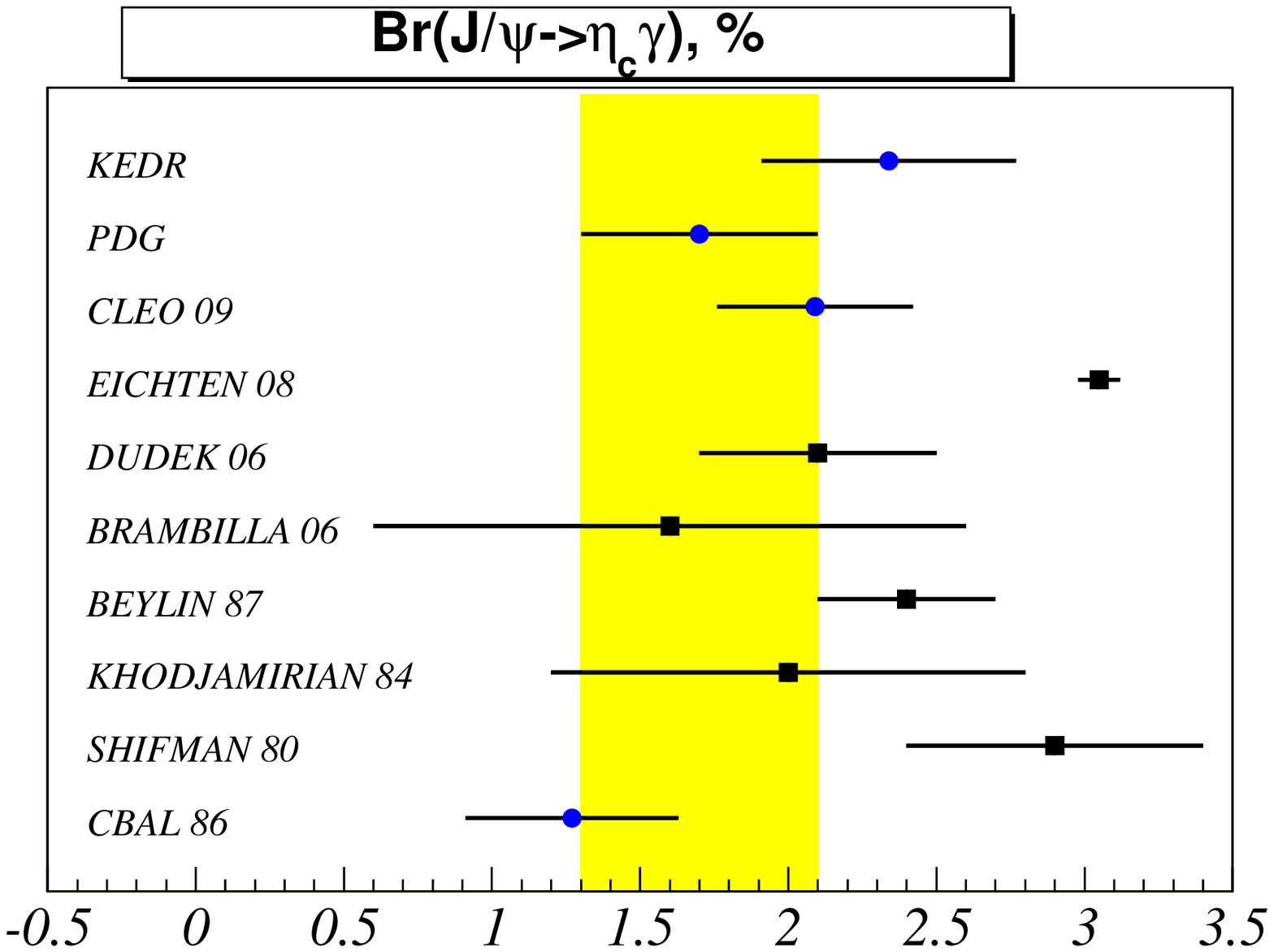}
\figcaption{\label{fig7} $\eta_{\rm c}$ mass, width and  ${\cal B}(J/\psi\to\eta_{\rm c}\gamma$) measurements. 
The theoretical predictions for the branching fraction are also shown (squares). 
A small error of Eichten 08 prediction is a consequence of the fact that the authors take into account 
the experimental uncertainty of the $J/\psi$ total width only and ignore all theoretical uncertainties.}
\end{center}

\begin{multicols}{2}

\end{multicols}

\end{document}